# DESIGN DIDN'T PREVENT THESE HAZARD EXPOSURES

David. E Mertz, P.E.
Senior Member, IEEE
Fermi Research Alliance, LLC
P.O. Box 500
Batavia, IL 60510
USA
d.e.mertz@ieee.org

*Abstract* – Two recent electrical incidents demonstrate how the design of equipment encouraged electrical workers to take actions that violated NFPA 70E principles. The features that encouraged non-compliant work execution will be described, as well as how the equipment was improved to facilitate safe work practices. Design-stage processes that help identify features that will foster rather than compromise safe work practices will be identified as well.

## I. INTRODUCTION

The design of electrical equipment is driven by many factors. The key factor has certainly always been the intended use of the equipment, and cost has never been too far behind. Early on, safety became another significant factor, and even a century ago it drove innovations that at the very least could be leveraged for market advantage. See figure 1.

Ideally this attention to detail would have so thoroughly pervaded electrical equipment design that Prevention through Design, also known as Safety by Design, principles are always consistently applied in the development of all electrical equipment. Too often, though, suboptimal equipment design contributes to putting workers in harm's way or encouraging the choice of less than adequate means and methods. Two examples of this contributed to instances of electrically inadequate work at Fermilab in 2024.

Fermilab is a Research and Development laboratory established by the U. S. Atomic Energy Commission in 1967 which continues its research mission today under the auspices of the U. S. Department of Energy. Among the lab's accomplishments are the first direct observations of the Bottom and Top quarks and the Tau neutrino. It continues in operation today with a primary mission to observe and quantify the many unusual properties of neutrinos, which may lead to a better understanding of fundamental particle physics. The electrical equipment at Fermilab ranges from the unique and exotic needed to accelerate protons to 99.999% of the speed of light, as well as much more familiar equipment that distributes power across the 6800 acre site, keeps the buildings warm and dry and keeps the building occupants comfortable.

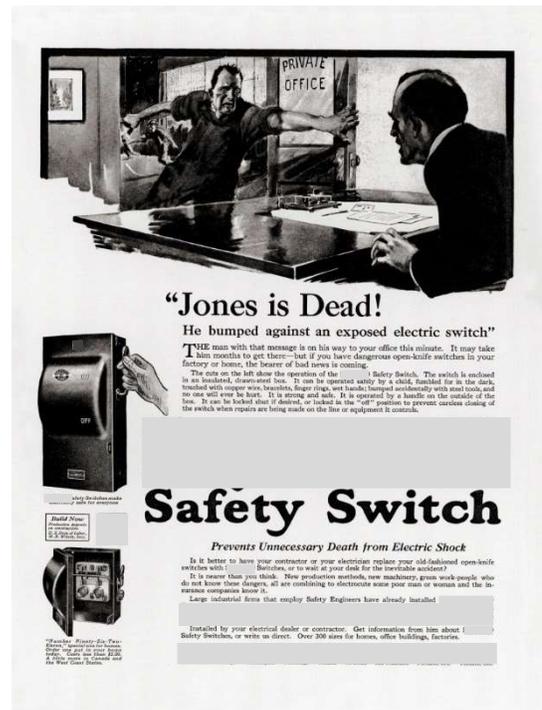

Figure 1: Safety Switch Advertisement

## II. EQUIPMENT EXAMPLES

A.  *The Legacy Design*

Among the ranks of exotic equipment is the 400 MeV Linear Accelerator, or Linac, which gives Fermilab's proton beam its first real burst of speed The protons are propelled by radio-frequency energy through a series of drift tubes in nine vacuum chambers, or "tanks." The Linac has been in regular operation since 1971, and with a few exceptions, most of the original equipment remains in service today.

The radio-frequency energy for each of the nine tanks come from a radio-frequency amplifiers based on designs used for broadcast applications. The amplifiers' accommodations for worker safety are typical of the era in which they were built.

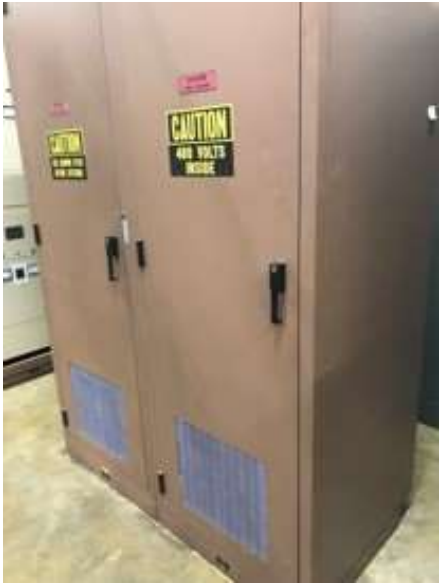
*Figure 2: Crowbar Cabinet*

Each amplifier has a crowbar cabinet, see figure 2, The crowbar cabinet detects arcs in the modulator that drives the RF power amplifier and triggers a mercury-based ignitron switch that instantly shorts the modulator capacitor bank to ground to avoid damaging the power amplifier tube and inhibits further operation of that RF section.

Located in the bottom center of this crowbar cabinet is a smaller enclosure known as the crowbar assembly. This assembly contains two 120 VAC to low voltage DC power supplies and the control logic that triggers the ignitron firing circuit which draws current from a 120 VAC to 385 VDC power supply it also contains. The location of this assembly is shown in figures 3 and 4.

The 120 VAC power to each crowbar assembly is supplied by a control power circuit with a 20-ampere circuit breaker that serves several other control-type loads at its particular RF amplifier. Age has made several of the other loads on these control circuits susceptible to failure when the power to them is cycled, so several years prior, the Linac electrical maintenance team (LEMT) replaced an external terminal strips on the crowbar assemblies with MS-type connectors, allowing the power to be removed from the assemblies under the cord and plug exception in OSHA 1910.147(a)(2)(iii)(A) without interrupting power to the other failure-prone loads.

On the day of the incident, the LEMT replaced a failed power amplifier driver tube at Linac Radio Frequency Station 3 (LRF3). Upon completion of this task, LRF3 was still not operating correctly and diagnostics with a good driver tube in place indicated that the crowbar assembly had also failed. A member of that team proceeded to remove the crowbar assembly so a spare assembly could be installed.

As can be seen from Figures 3 and 4, an installed crowbar assembly is in an ergonomically challenging location. It is not only located at the very bottom of the crowbar cabinet, but the fixed vertical column between the two doors is directly in front of its MS connector.

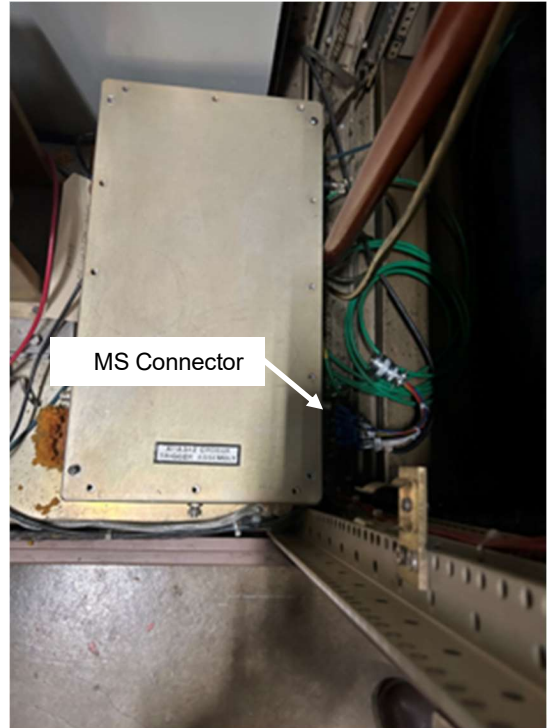
*Figure 3: Bottom Left Interior of Crowbar Cabinet*

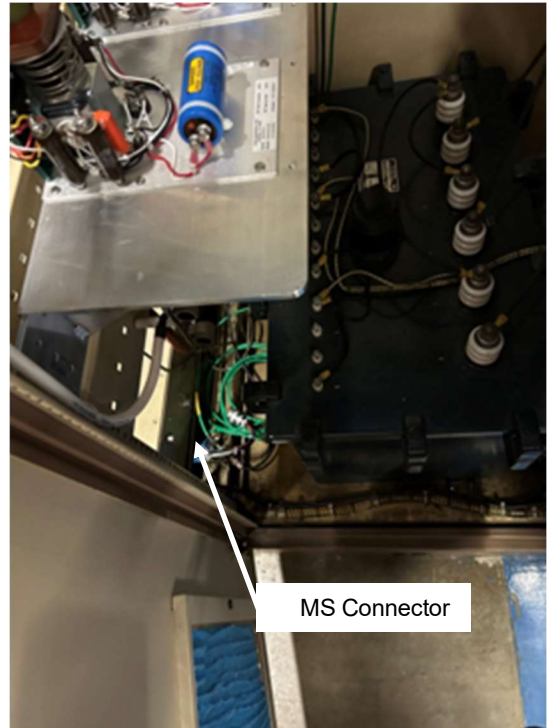
*Figure 4: Bottom Right Interior of Crowbar Cabinet*

The LEMT member found it impossible to reach the connector and still be able to exert the force needed to twist its locking ring, so he decided to shift the crowbar assembly to the left to better access the connector. To shift the crowbar assembly over, its mounting bolt needed to be removed first. Figure 5 shows the location of that bolt:

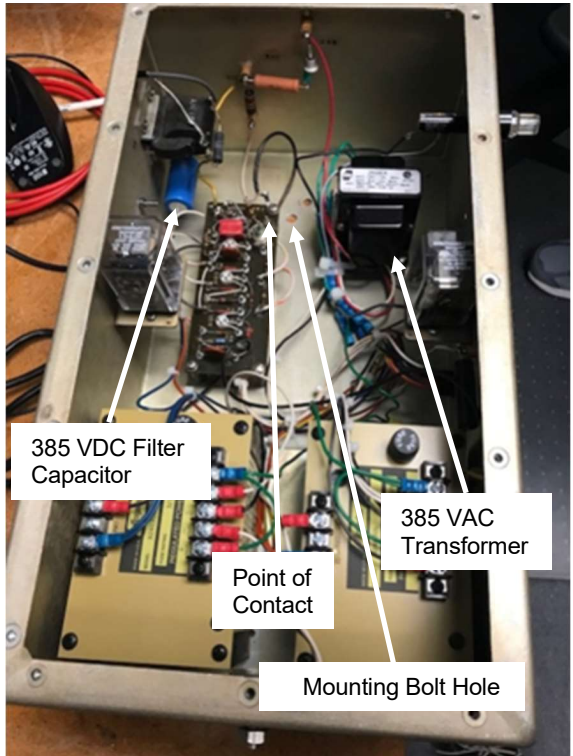

Figure 5: Crowbar Assembly Interior

The crowbar assembly mounting hole is indeed located inside the assembly enclosure, so the LEMT member removed the cover and proceeded to remove the bolt. Prior practice in the Linac had not identified any hazardous energy inside the crowbar assembly that would require additional protective measures. As the LEMT member removed the bolt, hand contact was made with a point on the printed circuit board energized at 385 VDC. The LEMT member did not immediately recognize the sensation as a shock and proceeded to shift the crowbar assembly and unplug the MS connector. After completing the replacement, the LEMT member mentioned that he might have received a DC shock and the event investigation was initiated.

The transformer producing the 385 VAC is rated 25 watts, for a full load amperage of 63 mA, which exceeds the 40 mA threshold in NFPA 70E 350.9(2), and the 20 µF filter capacitor stores 1.5 Joules of energy, which exceeds the 1 Joule threshold in NFPA 70E 350.(9)(3)(b). No injury occurred, and there was no noticeable mark at the point of contact with the LEMT member's hand.

B. The Modern Design

Suboptimal equipment design is not relegated to history. A second incident occurred about a month before the LRF3 event. While the LRF3 crowbar assembly is, if not unique, at least not ordinary, this one involved a very common piece of equipment – a sump pump controller located in Fermilab's Main Injector Service Building 40 (MI-40). When the Main Injector was constructed in the early 1990s, a simplex (one pump) sump pump was installed in the beam tunnel there. Because this tunnel is considered a radiation area, access is prohibited when the proton beam is operating in the Main Injector, and at other times it is restricted to minimize exposure to residual radiation. For over two decades any failures of this simplex controller or its pump required a quick shutdown of the proton beam and repair work in an area that was potentially more radiologically active than optimal.

To mitigate this problem, replacements of the MI-40 simplex sump pump and a similar one at the MI-62 service building with duplex (two pump) sump pumps were planned for the annual summer shutdown in 2020. The duplex controller shown in Figures 6 and 7 was purchased and installed at MI-40.

This duplex controller was built by one firm in 2019 and was purchased and relabeled by a second firm, which then sold it to Fermilab in the summer of 2020, which was at the height of the COVID-19 pandemic. The lapse of time and turnover of engineering and procurement personnel has made it difficult to determine when, why, and by whom certain decisions were made during the acquisition process, but this duplex controller did not conform in several ways to the specifications in the Request for Proposals. It is likely that supply chain constraints from the pandemic were a factor in those decisions.

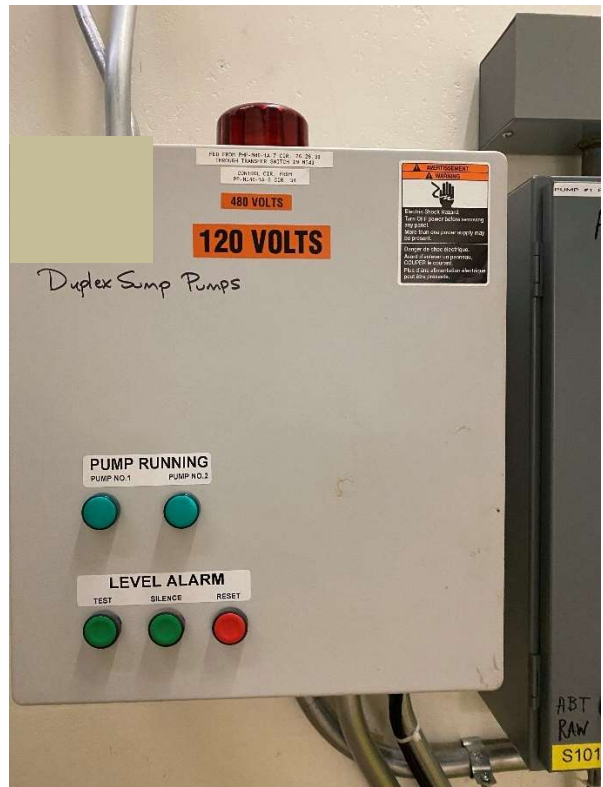

Figure 6: MI-40 Duplex Controller Exterior

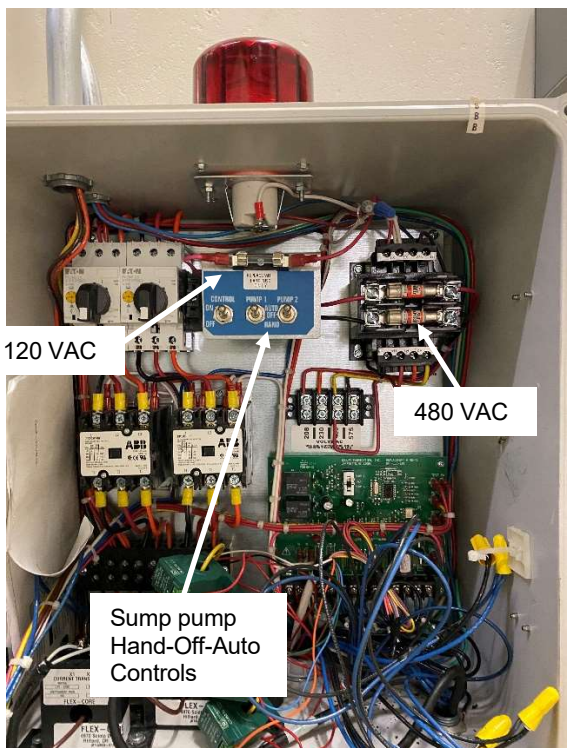
*Figure 7: MI-40 Duplex Controller Interior*

When this incident occurred, failure of this duplex sump pump had caused minor flooding of the MI-40 beam tunnel, which shut down proton beam operation. Measures not relevant to this subject were taken to address possible radiological issues with the water in the tunnel, which once completed, allowed three facility electricians to enter the tunnel with mechanics and radiological staff to determine why the duplex sumps weren't functioning correctly. Because of the radiological status of the area, the entry team wore protective clothing made of HDPE fibers but did not bring electrical PPE with them to avoid the risk of radiologically contaminating it.

Once the entry team had reached the duplex pumps, a visual inspection confirmed that the water level had come nowhere close to any vulnerable equipment. The obvious first step in troubleshooting a sump pump system is to run the pumps manually. With few exceptions, all sump pump controllers at Fermilab have exterior Hand-Off-Auto (HOA) pump controls that permit the basic functioning of the pump(s) to be tested without any exposure to electrical hazards. These exterior controls are required by the lab's standard pump controller specifications.

The lead electrician, E1, had more than a decade of experience at the lab and led the troubleshooting. E1 opened the duplex controller without performing LOTO or donning electrical hazard PPE. Faced with the HOA controls inside the cabinet, he asked the two other electricians, E2 and E3, both of whom had only a few months of experience at the lab, if they thought it was safe to reach in and operate the HOA switches. With their concurrence, E1 operated the switches. Once those tests were done, E1 closed the panel and completed the work to correct the problem, which was tangled cords for the sump level sensing floats. After the entry team exited the beamline enclosure, one of the radiological support team members described the incident to a member of the lab's general safety staff. Based on inspection of a spare duplex controller identical to the ones for MI-40 and MI-62, it was determined that E1 had been within the Restricted Approach Boundary for 480 volts without proper electrical hazard PPE, so an event review was initiated.

### III. INCIDENT REVIEW RESULTS

#### A. Linac LRF3 Crowbar Assembly

During the 2024 summer shutdown, all six RF stations that use this crowbar assembly and the nine crowbar assemblies (six in service, three spare) were retrofitted to no longer require the internal mounting bolt. A grounding strap was added in each RF station and an external grounding stud was added to each crowbar assembly to provide any grounding connection that the mounting bolt might have previously provided. Inspections were performed and discussions with LEMT members were held to identify any similar ergonomic or mounting issues, but none were found.

More extensive renovations and upgrades were not considered because the existing Linac will be soon replaced by a new superconducting Linac, which is being presently constructed by the Proton Improvement Plan II (PIP-II) project. The results of this incident review were shared with the PIP-II project, other facilities at Fermilab that use RF equipment, and with other Department of Energy Sites.

#### B. MI-40 Duplex Sump Pump Controller

The inspection of the identical spare duplex controller found that the internal HOA switches have a momentary, spring return action for the Hand position. With a momentary action on that switch, it would not be possible to place the controller in LOTO, operate the switch to the Hand position, and then remove LOTO and return power to the controller to check pump function. Plans were developed and materials purchased to retrofit these two controllers and the spare with external HOA switches. This was performed during a regularly-scheduled maintenance shutdown of the Main Injector. See Figure 8. Four other sump pump controllers without external controls have retrofits planned.

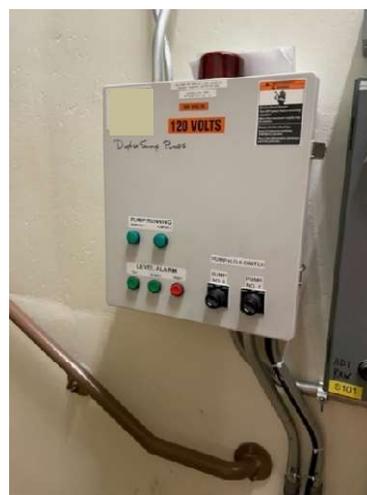
*Figure 8: Retrofitted Mi-40 Duplex Controller*

There are many locations, such as municipal lift stations in public areas, where unhindered access to system controls is unwise. There are serval common ways to restrict access to system controls without unnecessarily exposing workers to electrical hazards:
- Key-operated external switches
- Lockable blister boxes or "speakeasy" doors with systems controls beneath them
- Double-door panels with lockable exterior doors, behind which are located interior door panels with the controls and indicators mounted on them that block exposure to potentially energized conductors or circuit parts.

A review of the catalog of the original manufacturer revealed that they did offer sump pump controllers with external HOA controls, such as the one in Figure 9 that would have been completely suitable for the MI-40 application:

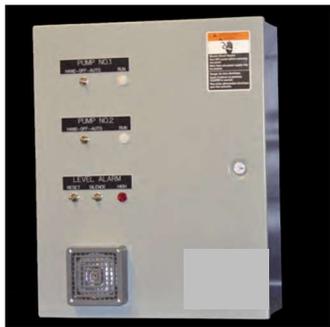

*Figure 9: Duplex Controller with External HOA*

This firm "…saw the need for quality cost effective control panels, suitable for almost any environment. The Series 2 controls were designed to fill that need." Figure 6 shows the Series 2 controller that was installed at MI-40. While it is true that with the proper electrical hazard PPE these panels could be used without an uncontrolled exposure to hazardous energy, there remains a significant difference between even a "controlled" exposure to hazardous energy and no exposure to it at all. The Hierarchy of Controls in NFPA 70E 110.1(H)(3) make it clear that the Engineering Control of the closed control panel is preferable to relying on personal protective equipment.

## IV. WORK PLANNING AND CONTROL

A key element for protecting workers from hazards of any sort is thorough work planning and control. In both of the instances described, the work planning was inadequate. In the first case, the original work scope was well planned, but when the additional scope of replacing the crowbar assembly was identified, no pause was taken to identify the hazards and implement mitigations. In the second case the work planning deficiencies were more systemic and beyond the scope of this paper. No procedures or steps were developed to direct the workers' activities once they had accessed the sump area, which troubleshooting activities were permitted and which weren't, or to identify at what point work should stop.

## V. CONCLUSIONS

While it is difficult to imagine a situation in which the design or condition of equipment can be the sole reason for an uncontrolled exposure to hazardous energy, it certainly can, as these two events demonstrate, make it more difficult to perform work in an electrically safe manner.

NFPA 70E Informative Annex O, *Safety-Related Design Requirements,* recommends in O2.2 that "design option decisions should facilitate the ability to eliminate hazards or reduce the risk…." Design decisions can readily impact the ability of many people to perform work in both a safe and an efficacious manner. As illustrated by the second example, design decisions are not only made by equipment manufacturers, but are also made by those who select the equipment that will be included in the design of specific facilities. Design decisions that place obstacles to the safe performance of work are effectively decisions to encourage work to be performed unsafely.

The electrical industry since its infancy has sought ways to protect both the public and workers in the electrical industry from the hazards electricity poses. Many product designs have advanced this cause, but until they are actually installed and used, they will not have any beneficial effect. In some instances it may take codes and regulations to drive implementation, but codes and regulations remain minimum requirements. The Hierarchy of Control will never be as prescriptive as other parts of our codes and standards, but it is hard to overstate how its principles can help design professionals focus on how their design choices will affect the people who interact with the results.

## VII. VITA

David E. Mertz, P. E., (S '82, M '89, SM '99) is the Electrical Safety Officer for the U. S. Department of Energy's Fermi National Accelerator Laboratory in Batavia, Illinois, USA. His electrical career began as a teen, troubleshooting and repairing TVs, stereos, and other home electronics. He then applied those skills in industrial environments while completing his B.S.E.E. at Valparaiso University. He continued working for Inland Steel in industrial automation and metallurgical research after graduation, later investing two decades providing consulting engineering services to heavy industrial, transportation, institutional, semiconductor, pharmaceutical, and R&D clients, and served in various officer roles for the Chicago Chapters of IEEE's Industry Applications and Power Engineering Societies. Eleven years ago he became Fermilab's Electrical Safety Officer, responsible for the lab's Electrical Safety Program.